# Tool steel ion beam assisted nitrocarburization

L.F. Zagonel*, F. Alvarez

*Instituto de Física "Gleb Wataghin", Universidade Estadual de Campinas, Unicamp,
13083-970 Campinas, São Paulo, Brazil*

## Abstract

The nitrocarburization of the AISI-H13 tool steel by ion beam assisted deposition is reported. In this technique, a carbon film is continuously deposited over the sample by the ion beam sputtering of a carbon target while a second ion source is used to bombard the sample with low energy nitrogen ions. The results show that the presence of carbon has an important impact on the crystalline and microstructural properties of the material without modification of the case depth.

### 1. Introduction

Nitrocarburizing is a thermochemical treatment applied mainly to iron alloys to improve fatigue strength, wear resistance, and corrosion resistance [1,2]. These improvements are attained by the diffusion of nitrogen and carbon beneath the surface, the formation of a compact compound layer (iron carbonitrides), and the formation of a thick diffusion zone, i.e., a region where the matrix nitrogen and carbon solubility limit is not exceeded. The key point is that, depending on the nitrogen and carbon concentration, the compound layer may be formed by $\varepsilon$-Fe3(N,C), $\gamma'$-Fe4(N,C), or $\theta$-Fe3C phases or a mixture of these phases. For practical applications, however, the formation of a $\varepsilon$-Fe3(N,C) monophase compound layer is preferable [3]. This is so because the crystalline parameters differ among phases induce inter-grain stress, weakening surface mechanical properties when the surface consists of several phases.

The processes for nitrogen and carbon incorporation into the steel normally use gases, salt bath, or plasma, among others [4,5]. Gas and salt bath are well established for industrial application. However, they present restrictions related to handling and environmental problems due to the use of explosive or toxic materials [6]. The plasma nitrocarburizing prevents some of these problems but, unlike from gas or salt bath, have not been reported to produce a monophase $\varepsilon$ compound layer [7]. Different approaches have been proposed and studied to produce better surface properties with nitrogen and carbon inclusion via plasma techniques, including pulsed plasma, r.f. plasma, and plasma immersion ion implantation [7–11]. These studies used plasmas composed of nitrogen gas (sometimes diluted in hydrogen gas) and, for carbon carrying gas, carbon dioxide, or some hydrocarbon gas such as methane or acetylene. The cited reports have in common the use of nitrogen and carbon ions with nearly the same energy and the presence of hydrogen (adding $H_2$ or as a constituent of the carbon carrying gas) in order to remove surface oxides that might prevent nitriding. Hydrogen is known to have effects on plasma, material surface, and species diffusion in the material bulk [12–14]. With regards to the ion energy, carbon was already reported to jeopardize nitrogen diffusion, making nitride layers thinner in plain steels as well as in austenitic stainless steels [8,11]. Indeed, the correct balance between carbon and nitrogen incorporation is the difficulty of mastering the nitrocarburized layer growth with the technique.

In this study the nitrocarburizing had been performed by ion beam assisted deposition (IBAD). Carbon atoms were sputtered from a target and deposited over the temperature-controlled sample while relative low energy nitrogen ions impinge the material surface. With this technique, the use of hydrogen is avoided and the observed effects concern carbon and nitrogen alone. The use of ion beam allows a fine control of the ion energy of the involved species, ion flux, and proportion of the species arriving at the substrate. The specific purpose of this work was two-fold:
(1) to analyze the effects on the microstructure of the amount of carbon at the surface sample during the ion beam nitriding, and
(2) to find the proper conditions leading to the formation of a $\varepsilon$ monophase compound nitride layer.

Table 1
AISI H13 steel composition, at.%

| Element | Content (±0.1) |
|---|---|
| Fe | 87.3 |
| C | 2.5 |
| Mn | 0.4 |
| Si | 2.1 |
| Cr | 5.9 |
| Mo | 0.8 |
| V | 1 |

## 2. Experimental

Rectangular samples (2mm×15mm×20 mm) were cut from a single tempered AISI-H13 lot (7.5±0.4 GPa bulk hardness). Table 1 displays the material composition as determined by chemical analysis. The slices were polished up to 5 mm diamond paste and cleaned in an acetone ultrasonic bath. One at a time, the samples were inserted into the vacuum system for IBAD and in situ X-ray photoemission spectroscopy (XPS). After this, the cross-sections etched with 5% nital were observed by field emission gun (FEG) and low vacuum (LV) scanning electron microscopes (SEM) (Jeol JMS-5900LV and JSM 6330F) at the Laboratório Nacional de Luz Síncrotron (LNLS Campinas-Brazil) facilities. The hardness profile of the samples was measured by nano-indentation on the mirror polished sample cross-sections. The data were analyzed by means of the Oliver–Pharr method and piling up effects were not considered [15]. The information regarding the crystalline structure was gathered in the Bragg–Brentano X-ray diffraction configuration using monocromatized Cu Kα radiation. Stress measurements were performed by the sin²Ψ method, and analyzed using the Poisson ratio and Young's modulus for this steel: 0.29 and 240 GPa, respectively.

The IBAD system is composed of two broad-beam Kaufman cells. One Kaufman cell (sputtering) is used for sputtering a carbon target while the other (assisting), pointing normally to sample surface, generates the nitrogen ion flux impinging the substrate. More experimental details are found elsewhere [16]. The energy of the sputtering cell was fixed at 1.45 kV and the current varied between 10 and 70 mA. These currents allow deposition rates of carbon over the sample, varying from 0.1 to 1.4 nm/min, as calibrated by deposition of carbon over a $Si_2O$ substrate. The current and energy of the assisting Kaufman cell were maintained at 30mA and 1.4 keV, respectively. In such configuration, the nitrogen beam erodes the surface by chemical and physical sputtering at rates as high as 2 mm/h. The nitrocarburizing was performed at a constant and relatively low temperature (400 _C) during 5 h for all the studied samples. Immediately after preparation, the sample is cooled to room temperature within a few minutes by moving the sample holder to a second holder maintained at room temperature.

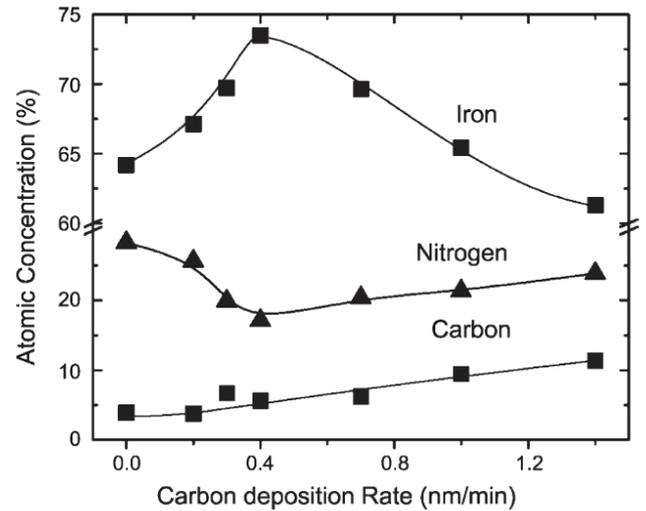

Fig. 1. Nitrogen, carbon, and iron surface concentrations for several carbon flux offers during nitrogen implantation. The lines are guide to the eyes.

## 3. Results and discussion

*3.1. Surface composition*

*In situ* XPS revealed the effect of carbon deposition on the surface composition during ion beam nitriding. As shown in Fig. 1, for small amounts of carbon supply, nitrogen content appreciably decreases. Nevertheless, above 0.4 nm/min of carbon supply, nitrogen and carbon surface concentration appear to increase linearly while iron content decreases. Chromium, the main alloying element of the sample, remains constant within experimental errors (not shown). Fig. 1 suggests that small carbon supply prevents nitrogen retention on the material surface. On the other hand, increasing the carbon supply, more nitrogen is retained on the material surface. The measured surface concentrations are compatible with $Fe_3(N,C)_{1+x}$ with $x$ varying from 0 to 0.5. It should be noted that XPS measured the first ~1 nm. Thus, the measured composition can be different beneath the surface. Finally, we note that oxygen concentration was negligible.

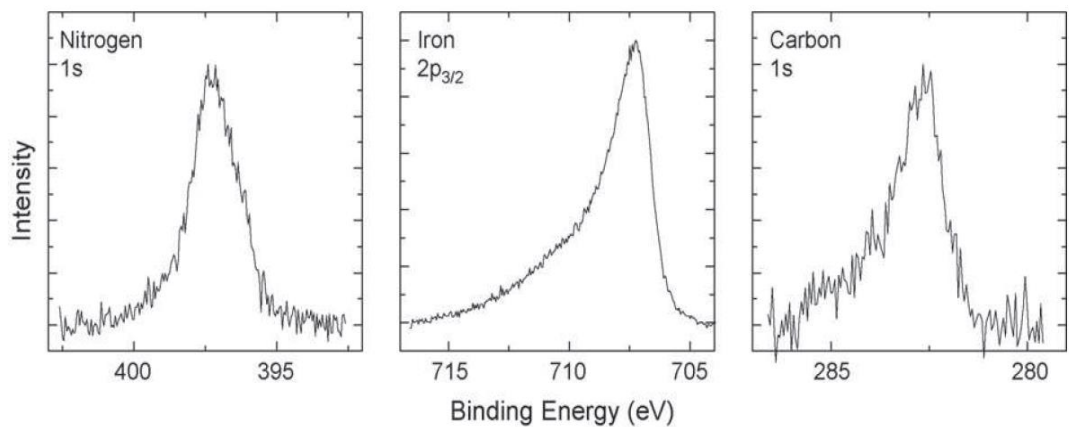

Fig. 2. Nitrogen, iron, and carbon photoemission electron core levels spectra for a sample nitrocarburized with a carbon flux offer of 1.4 nm/min.

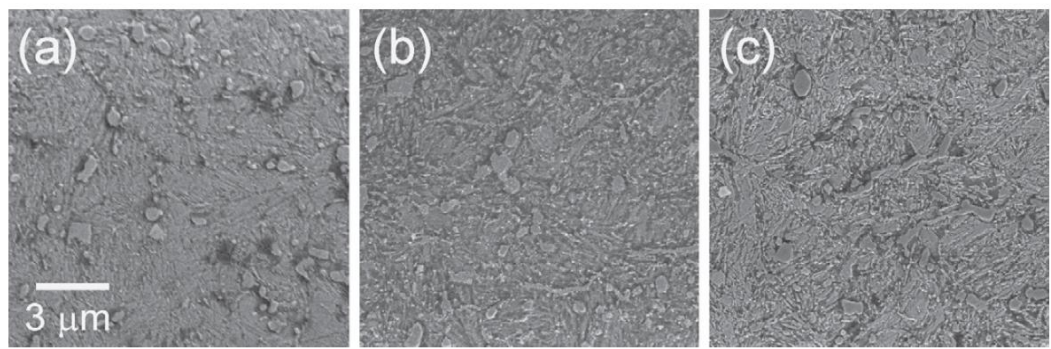

Fig. 3. Field emission gun scanning electron microscopy images: (a) nitrided sample without carbon flux offer; (b) nitrocarburized with carbon flux offer of 0.4 nm/min; (c) nitrocarburized with carbon flux offer of 1.4 nm/min.

The spectra of the photo-emitted core levels electrons corresponding to C 1s, N 1s, and Fe $2p_{3/2}$ are shown in Fig. 2. Regardless the carbon supply, the band associated with the C 1s core level always display a shoulder and a main peak centered at ~282.6 and ~283.8 eV, respectively. This indicates that carbon is likely to be in similar chemical environments for all the studied samples. Note that the physical and chemical sputtering induced by the assisting nitrogen ion beam is determining the actual surface composition measured by XPS. Unlike that the spectrum associated with the C 1s electron core level, the band associated with the N 1s electron core level depends on the carbon supply. Indeed, a slight change from 397.4 to 397.2 eV is observed going from no carbon supply to 0.4 nm/min carbon supply, respectively. The binding energies associated with N 1s and C 1s are lower than those reported for carbon nitride. Therefore, one can conclude that the surface is likely to be formed by iron–carbon–nitrides rather that carbonitride species [17,18].

### 3.2. Morphology

Fig. 3 shows the FEG-SEM pictures obtained in three studied samples: (a) without carbon supplies; (b) 0.4 nm/min; (c) 1.4 nm/min carbon supplies. All samples were treated with the same nitrogen ion flux. The images correspond to the samples cross-sections and the samples' surfaces are located a few micrometers above the images; i.e.: the images show the top of the diffusion layer. Nevertheless, the pictures clearly show the influence on the microstructure due to the presence of carbon. Indeed, the changes observed on the microstructure seem compatible with an increasing amount of carbon on the steel's matrix.

In order to resolve the precipitates and morphology, we have obtained images at higher electron beam energies. Fig. 4 shows a LV-SEM image in the backscatter mode of the sample nitrocarburized with 1.4 nm/min carbon supply. Unlike what was observed in the untreated samples (or in the matrix) where the precipitates are spherical, the supply of carbon induces needle shaped structures. As determined by energy dispersive spectroscopy (EDS), chromium rich precipitates appear dark while molybdenum rich precipitates appear bright in Fig. 4. The case depth observed is also in agreement with the hardness profile.

### 3.3. Hardness profiles

The introduction of nitrogen and carbon is known to increase the hardness of materials. The hardness profile is also known to be in close relation to carbon and nitrogen concentration profile. Fig. 5 shows the hardness profile for a selected group of the studied samples. As observed, the hardness profiles seem to be independent of carbon supply. This means that carbon is not blocking nitrogen diffusion. We note also that the observed hardness profiles are in agreement with the case depths observed on SEM micrographs.

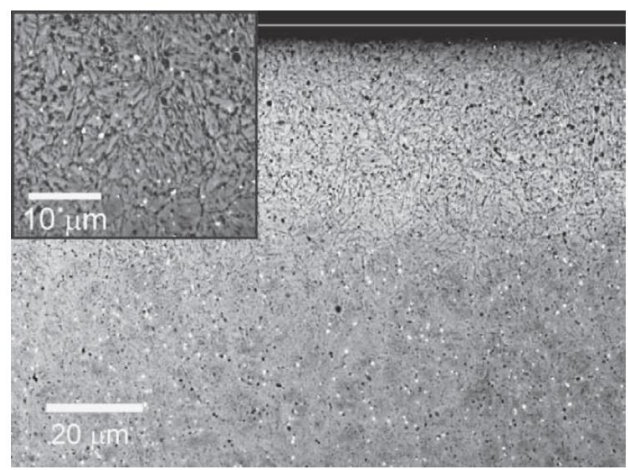

Fig. 4. Nitrocarburized sample with 1.4 nm/min carbon deposition rate. The presence of precipitates and a modified grain structure is observed (see inset). The white line above marks the real surface position.

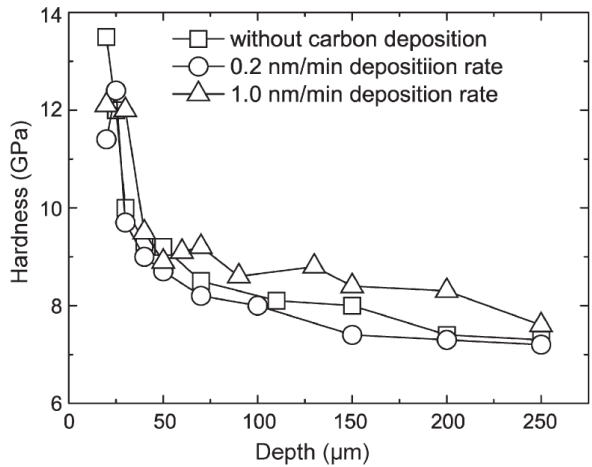

Fig. 5. Hardness profiles for selected samples. Others are omitted for sake of clarity.

*3.4. Crystalline structure of the compound layer*

Fig. 6 shows a summary from the X-ray diffraction results of nitrocarburized samples obtained at different treatment conditions together with an untreated sample for comparison. The diffractograms reveal the presence of a thin compound layer over the ferrite matrix. The analysis of the crystallographic phases provides a rough quantitative nitrogen and carbon content in the compound layer. First, the maximum nitrogen content is obtained without carbon supply, resulting in the formation of the $\varepsilon$-Fe$_3$N and $\gamma'$-Fe$_4$N phases (iron nitrides). Second, very low, 0.2 nm/min, carbon deposition rates cause a significant decrease in nitrogen content and just a slight amount of carbon incorporation (Fig. 1), resulting in a much smaller intensity on iron nitride/carbonitride phase peaks in the diffractogram (Fig. 6). Third, for carbon flux of 0.4 nm/min, the cementite phase seems to be also present. At last, for higher carbon deposition rates, 1.4 nm/min, the $\varepsilon$-Fe$_3$N(N,C) carbonitride becomes the main constituent of the thin compound layer. For all samples, the signal from the underneath ferrite phase is also observed.

The peaks associated with the a phase are broad and shifted to lower diffraction angles. The peaks broadening stems from two sources: (a) disorder, due to the presence of nitrogen and carbon in relative high concentrations and (b) the formation of precipitates. Finally, the peak shift is caused also by the tensile residual stress present under the compound layer. Indeed, for the samples treated with carbon fluxes higher than 0.4 nm/min, the residual stress is ~1.2 GPa. The calculated stress-free ferrite lattice space for these samples is ~0.2883 nm, i.e., 1% larger than the one observed in pure iron and the equivalent to those observed in untreated samples. The average region probed by X-ray diffraction is ~1 $\mu$m. Therefore, the presence of the peaks in the diffractograms associated with the $\alpha$ phase indicates that carbon and nitrogen concentrations in the compound layer as probed by XPS (first ~1 nm) is restrained in a thin zone.

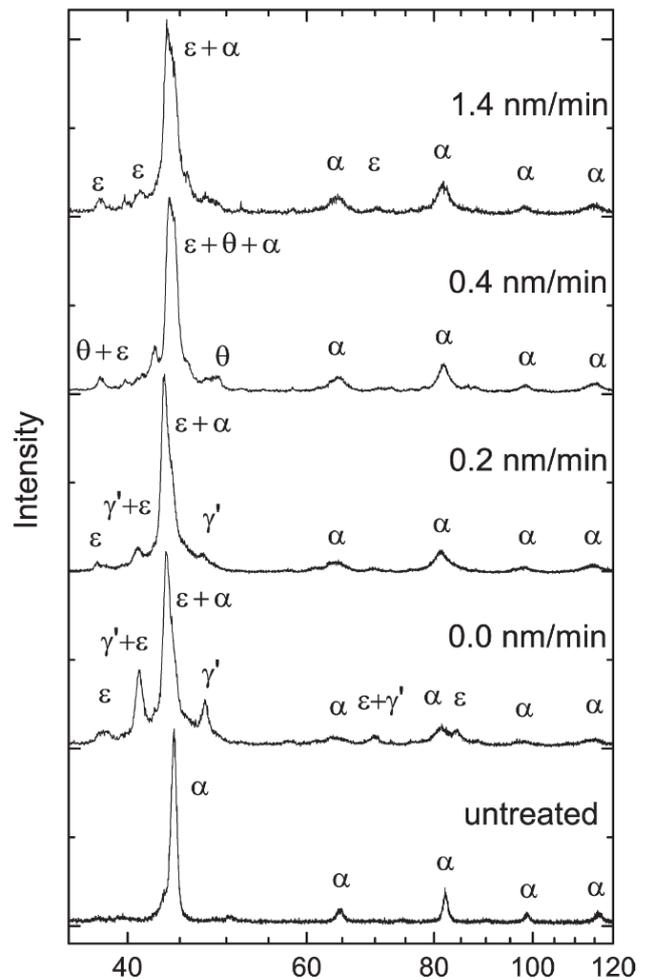

Fig. 6. Crystalline phases present on nitrocarburized samples with different carbon deposition rates together with an untreated sample for comparison.

**4. Conclusion**

Studies of ion beam assisted deposition as a technique for nitrocarburizing were presented. Surface composition revealed a two-fold behavior (carbon collaborates on high deposition rates but prevents nitrogen retention on low deposition rates). Therefore, carbon deposition has a great impact on the formation of nitrogen surface concentration even with the same nitrogen ion flux and energy. It has also been shown that nitrogen and carbon diffuse up to nearly 100 mm, with a ~40 mm layer presenting a hardness of ~12 GPa. Moreover, the carbon content on the surface can be accurately controlled by adjustment of carbon deposition rate, having a major impact on microstructure of the surface.

**Acknowledgements**

The authors are indebted to Prof. Eric Mittemeijer. Part of this work was supported from FAPESP project 02/12342-9. FA and LFZ are CNPq and Fapesp fellows, respectively. The authors would like to thank the LME/LNLS for technical support during electron microscopy work.